\documentclass[fleqn,10pt]{wlscirep}
\pdfoutput=1 
\title{Light-sound interconversion  in optomechanical Dirac materials}

\newcommand{\bk}[1]{\left(#1\right)}					
\newcommand{\Bk}[1]{\left[#1\right]}					
\newcommand{\bvec}[1]{\mathbf{#1}}
\newcommand{\absq}[1]{\left|#1\right|^2}				
\newcommand{\euler}[1]{\text{e}^{#1}}					

\usepackage{amsmath,amssymb,amstext}
\usepackage{braket}
\usepackage{ulem}

\author[1,*]{Christian Wurl}
\author[1,+]{Holger Fehske}

\affil[]{Institute of Physics, Ernst-Moritz-Arndt University Greifswald, Greifswald, 17489, Germany}
\affil[*]{wurl@physik.uni-greifswald.de}
\affil[+]{fehske@physik.uni-greifswald.de}

\begin{abstract}
Analyzing the scattering and conversion process between photons and phonons coupled 
via radiation pressure in a circular quantum dot  on a honeycomb array of optomechanical cells, we demonstrate 
the emergence of optomechanical Dirac physics. Specifically we  prove the formation of polaritonic 
quasi-bound states inside the dot, and angle-dependent Klein tunneling of light and emission of sound, 
depending on the energy of the incident photon, the photon-phonon interaction strength, and the radius of the dot.
We furthermore 
demonstrate that forward scattering of light or sound can almost switched off by an optically tuned Fano 
resonance; thereby the  system may act as  an optomechanical 
translator in a future photon-phonon
based circuitry.
\end{abstract}
\begin{document}

\flushbottom
\maketitle
%
%
\thispagestyle{empty}


The rapidly emerging field of optomechanics, describing the mechanical effects of light, opens new prospects for exploring hybrid quantum-classical systems which raise fundamental questions concerning the interaction and entanglement between microscopic and macroscopic objects~\cite{Viea07,LHM10,GKPBLS14},   classical-optical communication in the course of quantum information processing and storage~\cite{PHTSL13,SP11,FMLP16}, cooling of nanomechanical oscillators into their quantum ground state~\cite{Chea11,Teea11,FGN16}, or the development of nonclassical correlations~\cite{Riea16}, nonlinear dynamics, dynamical multistabilities and chaos~\cite{MHG06,QCHM12,WAF16,BAF15,CABF16};      
for a recent review see~\cite{AKM14}. 

Going beyond the prototyp cavity-optomechanical system consisting of a Fabry-Perot cavity with a movable end mirror, the currently 
most promising platforms are optomechanical crystals or arrays~\cite{ZS15,ECCVP09,SP10,SH14,LM13,HLQKM11}. These systems are engineered to co-localize and couple high-frequency (200-THz) photons and low-frequency (2-GHz) phonons. The simultaneous confinement of optical and mechanical modes in a periodic structure greatly enhances the light-matter interaction. Then the next logical step would be the creation of `optomechanical metamaterials' with an {\it in situ} tunable band structure,  which---if adequately designed---should allow to mimic classical dynamical gauge fields~\cite{WM16}, Dirac physics~\cite{SPM15},  optomechanical magnetic fields~\cite{SK15}, or topological phases of light and sound~\cite{PBSM15}, just as optical lattices filled with ultracold quantum gases ~\cite{Bl05} and topological photonic crystals~\cite{LJS14}.  Because of the ease of optical excitation, photon-phonon interaction control (i.e., functionalization)  and readout, artificial optomechanical  structures should be promising  building blocks of hybrid photon-phonon signal processing network architectures.   Thereby the complimentary nature of photons and phonons regarding their  interaction with the environment and their ability to transmit information over some distance will be of particular interest~\cite{SP11}. 

Here, we study a basic transport phenomenon in planar optomechanical metamaterials, the phonon-affected photon  transmission (reflection) through (by) a circular barrier, acting as a `qantum dot', created optically on a honeycomb  lattice.  Figure~\ref{fig.1} shows the `optomechanical graphene'  setup under consideration. Solving the scattering problem for a plane photon wave injected by a probe laser, we discuss Dirac polariton formation, possible Klein tunneling and photon-phonon conversion triggered by the (barrier-laser) tunable interaction between the co-localized optical and mechanical modes in the quantum dot region. The scattering of a perpendicularly incident (plane) photon wave by a planar barrier has been investigated with a focus on Klein-tunneling~\cite{SPM15}. Hence, to some degree, the present work can be understood as an extension of this study to the more complex quantum dot-array geometry, yielding a much richer angle-dependent scattering  and photon-phonon conversion. 

\section{Theoretical modelling}
\begin{figure}[t]
\center
\includegraphics[scale=0.3]{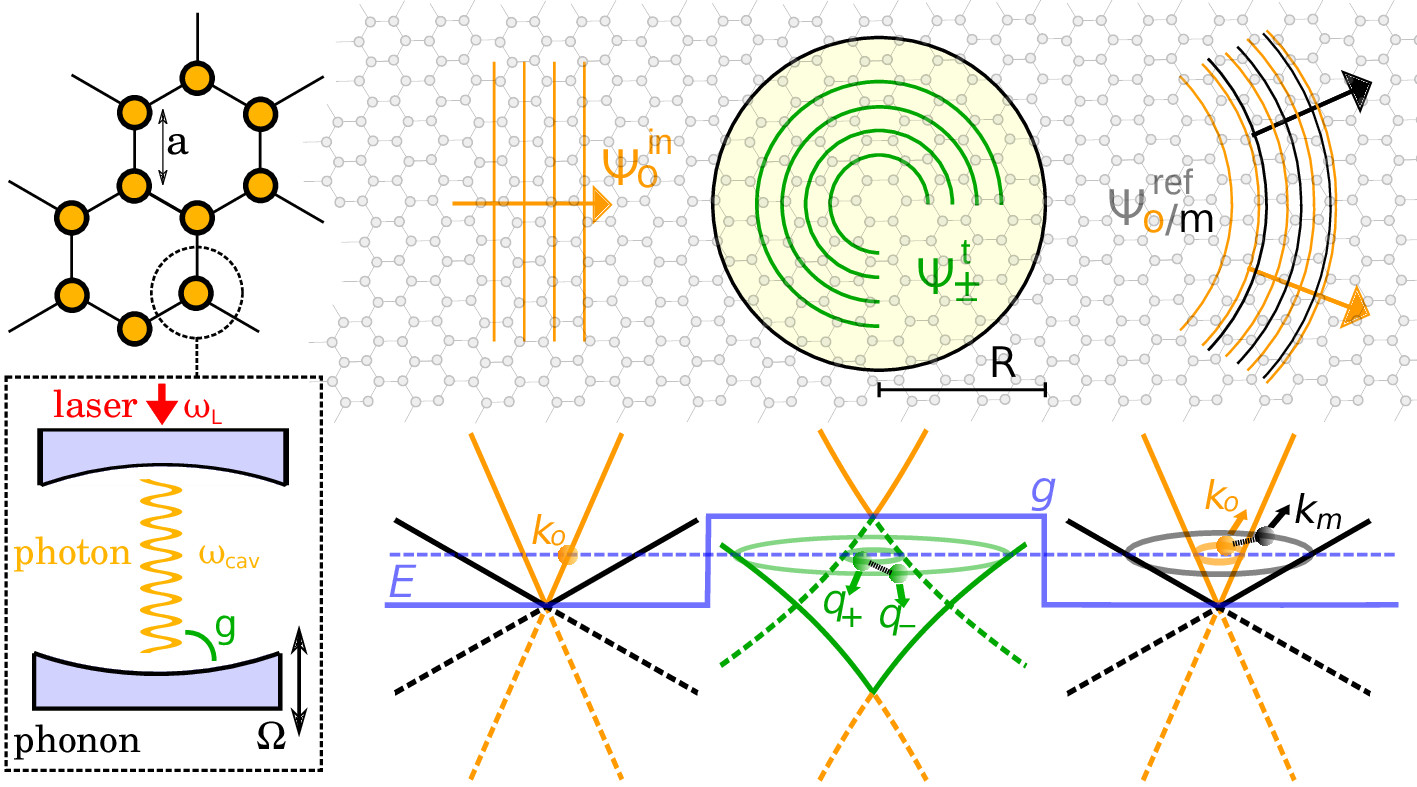}
\caption{(Color online)  Setup considered in this work. Left part: Optomechanical graphene. Honeycomb array of optomechanical cells driven by a laser with frequency $\omega_{L}$. The co-localized cavity photon ($\omega_{cav}$) and  phonon ($\Omega$) modes  interact (linearly) via radiation pressure  tunable by the laser power~\cite{AKM14}.   Upper right part: Scattering geometry. An incident optical wave ($\psi_o^{in}$, energy $E$, wavevector $\mathbf{k}_o \parallel \mathbf{e}_x$) hits the quantum dot  (radius $R$, photon-phonon coupling $g$); as a result transmitted polaritonic ($\psi^{t}=\psi_{+}^{t}+\psi_{-}^{t}$) and reflected ($\psi^{ref}=\psi_{o}^{ref}+\psi_{m}^{ref}$) waves appear (with wavevectors $\mathbf{q}_{\pm}$ and $\mathbf{k}_{o/m}$), which---due to the symmetry of the problem---carry an angular momentum, i.e., their wavevectors have components in any planar direction~\cite{HBF13,SHF15}.  Lower right part: Schematic bandstructure.  Without photon-phonon coupling the photon (orange) and phonon (black) Dirac cones (obtained in low-energy approximation) simply intersect.  In the quantum dot region with $g>0$, weakly non-linear (photon-phonon) polariton bands (green) emerge. Here, solid (dashed) lines correspond to pseudospin $\sigma=1$ ($-1$). Connecting lines between $\mathbf{q}_+$ and $\mathbf{q}_-$ ($\mathbf{k}_o$ and $\mathbf{k}_m$) indicate that the corresponding states are superimposed. The dashed (solid) blue line gives the energy $E$ (position-dependent profile of $g$). Model parameters: The continuum approximation is justified if $k \ll 1/a$ and $R \gg a$. Moreover, we have to avoid any `phonon lasing' instabilities, i.e., the photon transfer element $2v_{o}/3a$ has to  be smaller than $\Omega/3$~\cite{SPM15}. If so, the effects discussed in this paper should be experimentally accessible for $g/\Omega \ll 1$. With a lattice constant $a\sim 50 \mu \text{m}$~\cite{SP10}, a photon [phonon] transfer element $\sim \Omega/6$ [$\Omega/60$], and a membrane eigenfrequency $\Omega=-\Delta\sim 10 \text{MHz}$~\cite{SPM15}, the photons [phonons] group velocity $v_{o}$ [$v_{m}$] is about $10^3 \text{m/s}$ [$10^2 \text{m/s}$], and the optomechanical coupling $g$ should not exceed $0.1 \text{MHz}$. Then, $R \sim100a$.
\label{fig.1}}
\end{figure}

To formulate the scattering problem we follow the standard approach of (i) linearizing the dynamics around the steady-state solution within the 
rotating-wave approximation in the red-detuned ($\Delta=\omega_L-\omega_{cav}<0$) moderate-driving regime~\cite{AKM14}  and (ii) adapting the single-valley 
Dirac-Hamiltonian within the continuum approximation, valid for sufficiently low energies and barrier potentials that are smooth on the scale of the lattice constant $a$ but sharp on the scale of the de Broglie wavelength~\cite{HBF13}. Furthermore, focusing on the scattering by the barrier exclusively, we assume $\Delta=-\Omega$, and obtain (after the appropriate rescaling $H\to H/\hbar-\Omega$) the optomechanical Dirac-Hamiltomian~\cite{SPM15},
\begin{equation}\label{Hamiltonian}
H=\bk{\overline{v}+\frac{1}{2}\delta v \tau_{z}}\bvec{\sigma}\cdot \bvec{k}-g\Theta\bk{R-r}\tau_{x}\,,
\end{equation} 
as a starting point ($\hbar=1$).  Here, $\overline{v}=\frac{1}{2}\bk{v_{o}+v_{m}}$, $\delta v = v_{o}-v_{m}$, with $v_{o/m}$ as the velocities of the optical/mechanical 
modes, $\bvec{\tau}$ and $\bvec{\sigma}$ are vectors of Pauli matrices, $\mathbf{k}$ ($\mathbf{r}$) gives the wavevector (position vector) of the Dirac wave,  $R$ is the quantum-dot radius, and 
$g$ parametrizes the photon-phonon coupling strength, cf. fig.~\ref{fig.1}.   
The low-energy dispersion follows as
\begin{equation}\label{dispersion}
E_{\tau,\sigma} (\bvec{k})=\sigma \overline{v}\left| \bvec{k} \right|+\sigma\tau\sqrt{g^2+\frac{\delta v^{2}}{4}\left| \bvec{k} \right|^2}\,,
\end{equation}
where $\tau=\pm1$ denote the two-fold degenerate, non-linear polariton branches with  sublattice pseudospin $\sigma=\pm 1$. The eigenfunctions of~(\ref{Hamiltonian}) take the form
$\ket{\psi_{\tau,\sigma}}=\mathcal{N}_{\tau,\sigma}\ket{\sigma,\bvec{k}}\bk{g\ket{o}+\varepsilon_{\tau,\sigma}\ket{m}}$ 
with normalization $\mathcal{N}_{\tau,\sigma}=\bk{g^2+\varepsilon_{\tau,\sigma}^2}^{-1/2}$, $\varepsilon_{\tau,\sigma}=v_{o}\sigma k-E_{\tau,\sigma}$, and the bare (optical/mechanical) eigenstates $\ket{o/m}$ of $\tau_{z}$.  For $g=0$, the bandstructure simplifies to two independent photonic and phononic Dirac cones, and the scattering problem can be solved as for a graphene quantum dot~\cite{HBF13,SHF15,SHF15_2}.   

We expand the incident photonic wave (in $x$ direction),  the  transmitted wave inside the dot ($\psi^{t}=\psi^{t}_{+}+\psi^{t}_{-}$) and the reflected wave ($\psi^{ref}=\psi_{o}^{ref}+\psi_{m}^{ref}$) in polar coordinates ($l$ --  quantum number of angular momentum): 
\begin{align}\label{incwav}
\psi_{o}^{in}&=\frac{1}{\sqrt{2}}\euler{ik_{o}x} \binom{1}{1} \ket{o}=\sum \limits _{l=-\infty}^ {\infty} i^{l+1}\phi_{l}^{\bk{1}}\bk{k_{o}r} \ket{o},\\
\label{transwav}
\psi_{\pm}^{t}&=\mathcal{N}_{\pm}\sum \limits _{l} i^{l+1} t_{\pm,l} \phi_{l}^{\bk{1}}\bk{q_{\pm}r}\Bk{g\ket{o}+\varepsilon_{\pm}\ket{m}},\\
\label{refwav}
\psi_{o/m}^{ref}&=\sum \limits _{l}i^{l+1}   \sqrt{\frac{v_{o}}{v_{o/m}}} r_{o/m,l} \phi_{l}^{\bk{3}}\bk{k_{o/m} r} \ket{o/m}.
\end{align}
For  $E>0$, we can take $\sigma=+1$ and distinguish the branches of the incident and reflected waves by $\tau=\pm1$.  For the transmitted wave, where 
$\varepsilon_{\pm}=v_{o}\sigma_{\pm}q_{\pm}-E$, $E\gtrless g$ is possible  and we denote the two polaritonic branches   
by $+$ and $-$. Here, for $E>g$ ($E<g$) $\sigma_\pm=1$ ($\tau_\pm=-1$), and states with different $\tau_\pm=\pm1$ 
($\sigma_\pm=\pm1$) are superimposed, see fig.~\ref{fig.1}.
In eqs.~(\ref{incwav})-(\ref{refwav}) the eigenfunctions of the Dirac-Weyl Hamiltonian $\bvec{\sigma}\cdot\bvec{ k}$ are
\begin{equation}
\phi_{l}^{\bk{1,3}}\bk{kr}=\frac{1}{\sqrt{2}}\binom{-i \mathcal{Z}_{l}^{\bk{1,3}}\bk{kr}\euler{il\phi}}{\sigma \mathcal{Z}_{l+1}^{\bk{1,3}}\bk{kr}\euler{i\bk{l+1}\phi}}\,,
\end{equation}
where $\mathcal{Z}^{\bk{1}}_{l}=J_{l}$  [$\mathcal{Z}_{l}^{\bk{3}}=H_{l}^{\bk{1}}$]   are the Bessel [Hankel] function of the first kind (in the following we omit the upper index $\phantom{}^{\bk{1}}$ of the Hankel functions). The continuity conditions at $r = R$ give the reflection $r_{o/m,l}$ and transmission coefficients $t_{\pm,l}$:
\begin{equation}\label{eq_ref}
r_{o/m,l}=-\sqrt{\frac{v_{o/m}}{v_{o}}}\frac{Z_{o/m,l}}{\det A }, \quad t_{\pm,l}=-\frac{1}{\mathcal{N}_{\pm}}\frac{W_{\pm,l}}{\det A}.
\end{equation}
In eq.~\eqref{eq_ref}, $Z_{o,l}=\det A -igY$, and
\begin{align}\label{eq scph}
&Z_{m,l}= -i\varepsilon_{+}\varepsilon_{-}  \times \{ \bk{Y_{l}\bk{k_{o}R}J_{l+1}\bk{k_{o}R}-Y_{l+1}\bk{k_{o}R}J_{l}\bk{k_{o}R}} \times  \bk{\sigma_{+}J_{l}\bk{q_{-}R}J_{l+1}\bk{q_{+}R}-\sigma_{-}J_{l}\bk{q_{+}R}J_{l+1}\bk{q_{-}R}}\}, \\
\label{eq scm}
&W_{\pm,l}= \mp \varepsilon_{\mp} \times \{ \bk{H_{l}\bk{k_{o}R}J_{l+1}\bk{k_{o}R}-H_{l+1}\bk{k_{o}R}J_{l}\bk{k_{o}R}} \times\{   \bk{J_{l}\bk{q_{\mp}R}H_{l+1}\bk{k_{m}R}-\sigma_{\mp}J_{l+1}\bk{q_{\mp}R}H_{l}\bk{k_{m}R}} \},\\
\label{eq Y}
&Y=Y_{l}\bk{k_{o}R}  \times \{ \big [\varepsilon_{-}\sigma_{+}J_{l}\bk{q_{-}R}J_{l+1}\bk{q_{+}R} -\varepsilon_{+}\sigma_{-}J_{l}\bk{q_{+}R}J_{l+1}\bk{q_{-}R}\big] \cdot H_{l+1}\bk{k_{m}R} \nonumber \\
&+\sigma_{+}\sigma_{-}\bk{\varepsilon_{+}-\varepsilon_{-}}J_{l+1}\bk{q_{+}R}J_{l+1}\bk{q_{-}R}H_{l}\bk{k_{m}R}\} \nonumber \\
&+Y_{l+1}\bk{k_{o}R}  \times \{ \big [\varepsilon_{-}\sigma_{-}J_{l}\bk{q_{+}R}J_{l+1}\bk{q_{-}R} 
-\varepsilon_{+}\sigma_{+}J_{l}\bk{q_{-}R}J_{l+1}\bk{q_{+}R}\big] \cdot H_{l}\bk{k_{m}R} \nonumber \\
&+\bk{\varepsilon_{+}-\varepsilon_{-}}J_{l}\bk{q_{-}R}J_{l}\bk{q_{+}R}H_{l+1}\bk{k_{m}R} \}.
\end{align}
Here,  $\det A$ is obtained from eq.~\eqref{eq Y} when substituting $Y_{l\bk{+1}}$ by  $H_{l\bk{+1}}$ and multiplying by $g$. 
Note that the scattering coefficients are invariant under the transformation $\bk{E,g,R^{-1}} \rightarrow \bk{\gamma E,\gamma g,\gamma R^{-1}}$ with $\gamma \in \mathbb{R}$. Furthermore, the reflection coefficients have upper bounds:  $\left|r_{o,l}\right|\leq 1$ and $\left|r_{m,l}\right|\leq \sqrt{v_{o}/v_{m}}/2$.

From the current density of the reflected waves in the far field,
\begin{align}\label{j}
&j_{o/m}\bk{\phi} = \frac{4v_{o}}{\pi k_{o/m}r} \sum \limits _{l,l'=0}^{\infty}r_{o/m,l'}^{*}r_{o/m,l}   \\
&\times \Bk{\cos\bk{\bk{l+l'+1}\phi}+ \cos\bk{\bk{l-l'}\phi}} \nonumber, 
\end{align}
we obtain the scattering efficiency, that is, the scattering cross section divided by the geometric cross section, as
\begin{equation}\label{Q}
Q_{o/m}=\frac{4 }{k_{o/m}R} \sum \limits _{l=0} ^{\infty} \absq{r_{o/m,l}}.
\end{equation}
We note that in eqs.~\eqref{j}, \eqref{Q}, and hereafter, $l \geq 0$. The density $\rho=\psi^{\dagger} \psi$ and the current $\bvec{j}=\psi^{\dagger} \boldsymbol{\sigma} \psi$ in- and outside the quantum dot region further specify the scattering. 
\section{Numerical results and discussion}
Treating the scattering by  the circular quantum dot region numerically, we adopt $v_{m}=0.1v_{o}$ and employ units such that $v_{o}=1$. Moreover,  for the experimental reliable parameters quoted in the caption of fig.~\ref{fig.1}, fixing $g$, $100a$ is a natural unit for the quantum dot radius $R$, where the number of cells (defects) enclosed in the quantum dot region is about $10 ^{4} R ^{2}$. Due to the scale invariance of the scattering coefficients, in what follows all physical quantities will be discussed in dependence on $E/g$ and $Rg$. 
\begin{figure}
 \center
 \includegraphics[scale=0.261]{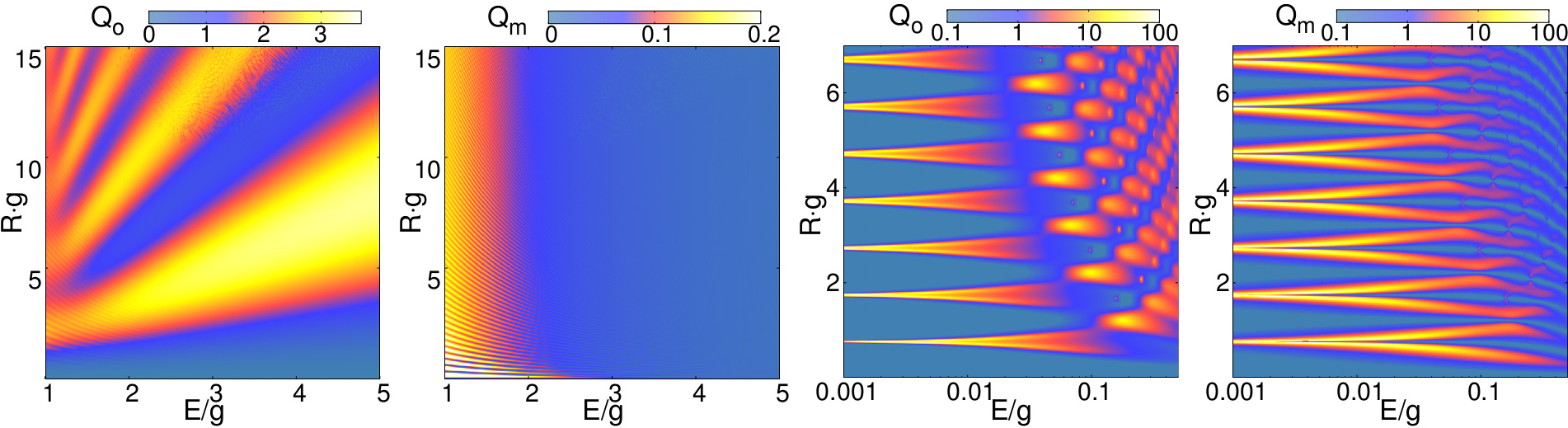}
    \caption{(Color online) Photonic/phononic scattering efficiency $Q_{o/m}$ in the $E/g$-$Rg$ plane. \label{fig.2}}
\end{figure}

Figure~\ref{fig.2} displays the complex pattern of both the photonic $Q_{o}$ and phononic $Q_{m}$ contributions to the scattering efficiency in the $E/g$--$Rg$ plane. When the photon hits the quantum dot it stimulates mechanical vibrations (phonons) because of the optomechanical interaction. Then both scattered waves are inherently correlated. For energies of  incident photon larger than the optomechanical coupling, $Q_o$ ($Q_m$) reveals a very broad (narrow) ripple structure with maxima of high (rather low) intensity. Above $E/g \sim 2$ the phonon is hardly scattered,  while the photon is still heavily influenced by the dot. This is because the phonon wave numbers take  large values very quickly, compared to those of the photon, simply because $v_m$ is smaller than $v_o$ by an order of magnitude. If the dispersion of the phonon is unaffected by $g$, the wave numbers inside and outside are almost identical and scattering disappears. The same, in principle, happens to the photon, but at much larger energies. In this limit, photon scattering resembles the scattering of ultrarelativistic Dirac particles, which are massless outside the dot and carry an effective mass  $m=g\sqrt{2/v_{o}^3\bk{v_{o}-v_{m}}}$ inside the quantum dot region (here, $v_{o}$ plays the role of vacuum ligth speed). 
\begin{figure}
\center
\includegraphics[scale=0.261]{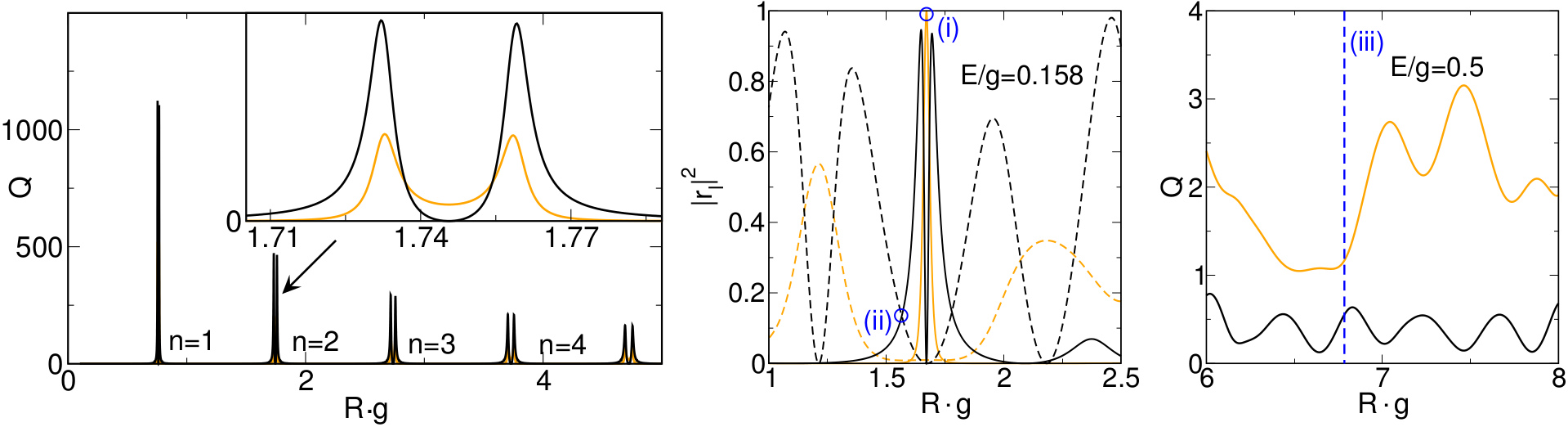}
\vspace*{-0.6cm}
\caption{(Color online) Left: scattering efficiency for photons (orange) and phonons (black)  in dependence on $Rg$. Here, $E/g=0.001$, i.e.,
the size-parameter $E R \ll 1$. For $n=2$, $Q_m$ vanishes at $Rg\simeq1.75$, whereas $Q_o$ stays finite (see inset). Middle: photonic (orange) and phononic (black) reflection coefficients with $l=0$ (dashed) and $l=1$ (solid) 
in dependence on $R g$, where $E/g=0.158$, i.e., the size-parameter $E R \lesssim 1$. For better comparison, the phononic coefficients 
were divided by their upper bound $v_{o}/4v_{m}$. Rigth: photonic (orange) and phononic (black) scattering 
efficiency at $E/g=0.5$; now $E R \gtrsim 1$. The cases $R g = 1.671$, $R g=1.566$ and $R g=6.78$ are marked by (i), (ii), and (iii), respectively.  \label{fig.3}}
\end{figure} 

The situation becomes much more involved when the energy of the incident optical wave is smaller than the optomechanical coupling, see the right panels in  fig.~\ref{fig.2} for $E/g<1$. Let us first consider the case where the size-parameter $ER$ is very small, i.e., the wavelengths $2\pi /k_{o/m}$ are large compared to the quantum dot radius $R$. In fig.~\ref{fig.2} this corresponds to the region 
$E/g \lesssim 0.01$. Here, sharp scattering resonances occur at a sequence of equidistant radii.  The left panel in fig.~\ref{fig.3} gives a closer look at this limiting behavior  and demonstrates that in each case two resonances occur, in fact, symmetrically around a point where the phonon scattering vanishes while the photon scattering is small but finite (see inset). These resonances, numbered by $n \in \mathbb{N}$, belong to the lowest photonic/phononic partial waves with $l=0$.  Expanding the phononic reflection coefficients~\eqref{eq scph} with respect to the small size-parameter $ER$, the phonon-scattering depletion points  result as $Rg=j_{l,n}\sqrt{v_{o}v_{m}}$, where $j_{l,n}$ are the $n$-th zero of the Bessel function $J_{l}$. We note that here the phonon resonance peaks are larger than the photonic ones.  Of course, such resonances also occur for the next higher partial wave with $l=1$ at $Rg=j_{1,n}\sqrt{v_{o}v_{m}}$, but are not visible in fig.~\ref{fig.3} left on account of their tiny linewidth/intensity.

In case that the size-parameter $ER \sim 1$, the wavelengths $2\pi /k_{o/m}$ are in the order of the  dot radius $R$. In this regime,
only the lowest partial waves will be excited to any appreciable extent, and the photonic [phononic] resonances appear as bright spots [splitted stripes] at specific `points' [lines] in the $E/g$-$Rg$ plane, see fig.~\ref{fig.2}.  The linewidths get smaller for larger  $l$, once one of the reflection coefficients $r_{o,l}$ ($r_{m,l}$) reaches unity (their upper bound). The photonic resonances with even (odd) $l$ are approximatively located at $Rg=j_{1(0),n}\sqrt{v_{o}v_{m}}$, where the phononic scattering is perfectly suppressed. This is illustrated by the middle panel in fig.~\ref{fig.3}: At $Rg \simeq 1.7$ [case (i)], the $l=1$ photon mode is resonant and the scattering becomes purely photonic (i.e., the contribution of all phonon modes goes to zero). The  phonon resonances of the $l=1$ mode appear symmetrically about  this photon resonance (at these points, on the other hand,  the photonic contribution is significantly weakened). A similar scenario arises for the resonance of the $l=0$ modes at $Rg \simeq 1.24$ and $Rg \simeq 2.24$. Vice versa, at certain radii the scattering becomes purely phononic, see, e.g., case (ii) where $Rg = 1.566$. This allows one to switch from entirely photon to phonon scattering just by varying  the dot radius. 

If the size-parameter increases further, the situation changes again. Now even higher partial waves will be excited. In this regime, the photon scattering efficiency is always a larger than the phononic one.  Approximating the resonance points by the zeros of the Bessel function is no longer possible; as a result both $Q_{o}, Q_{m}>0$, cf. fig.~\ref{fig.3} right. In the extreme limit $ER\gg 1$, however, phonon scattering is negligibly small and does not have to be considered.  
\begin{figure}
\center
\includegraphics[scale=0.264]{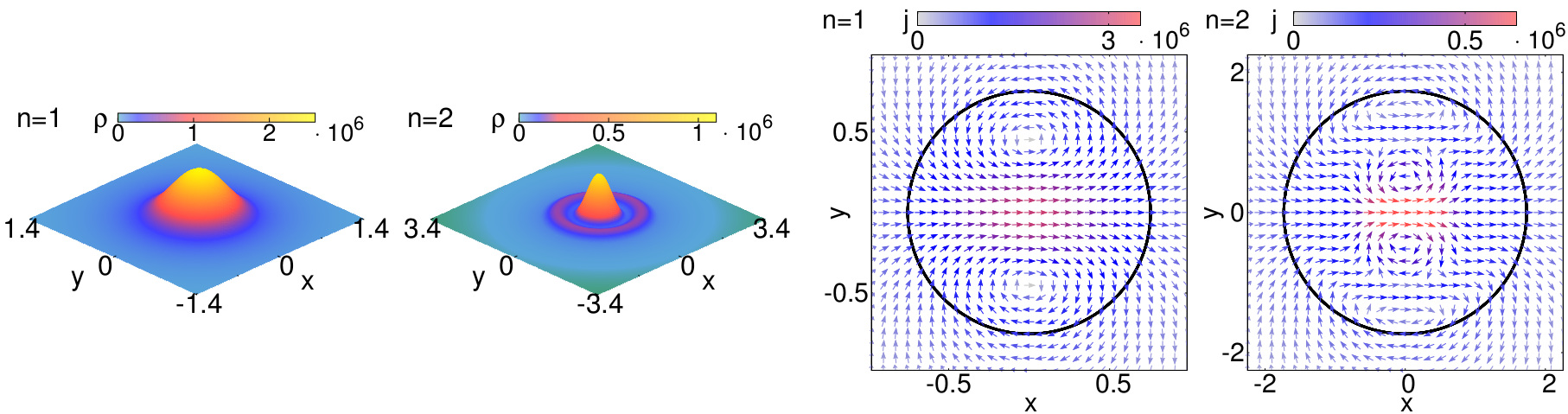}
\caption{(Color online) Scattering characteristics in the near field. Shown are the probability density $\rho=\psi^{\dagger}\psi$ (left) and the current density $\bvec{j}=\psi^{\dagger}\bvec{\sigma}\psi$ for $l=0$ (right; the circle marks the quantum dot), where $\psi=\psi^{t}$ inside and $\psi=\psi^{in}+\psi^{ref}$ outside the dot. 
Results correspond  to the resonances $n=1$ and $n=2$ given by fig.~\ref{fig.3} (left) and we have chosen $R=0.754$ for $n=1$ and $R=1.732$ for $n=2$ (with $g=1$), 
where $Q_o=Q_m$ (crossing of black and orange lines in the inset of the left panel in fig.~\ref{fig.3}). \label{fig.4}}
\end{figure}
\begin{figure}
 \center
 \includegraphics[scale=0.22]{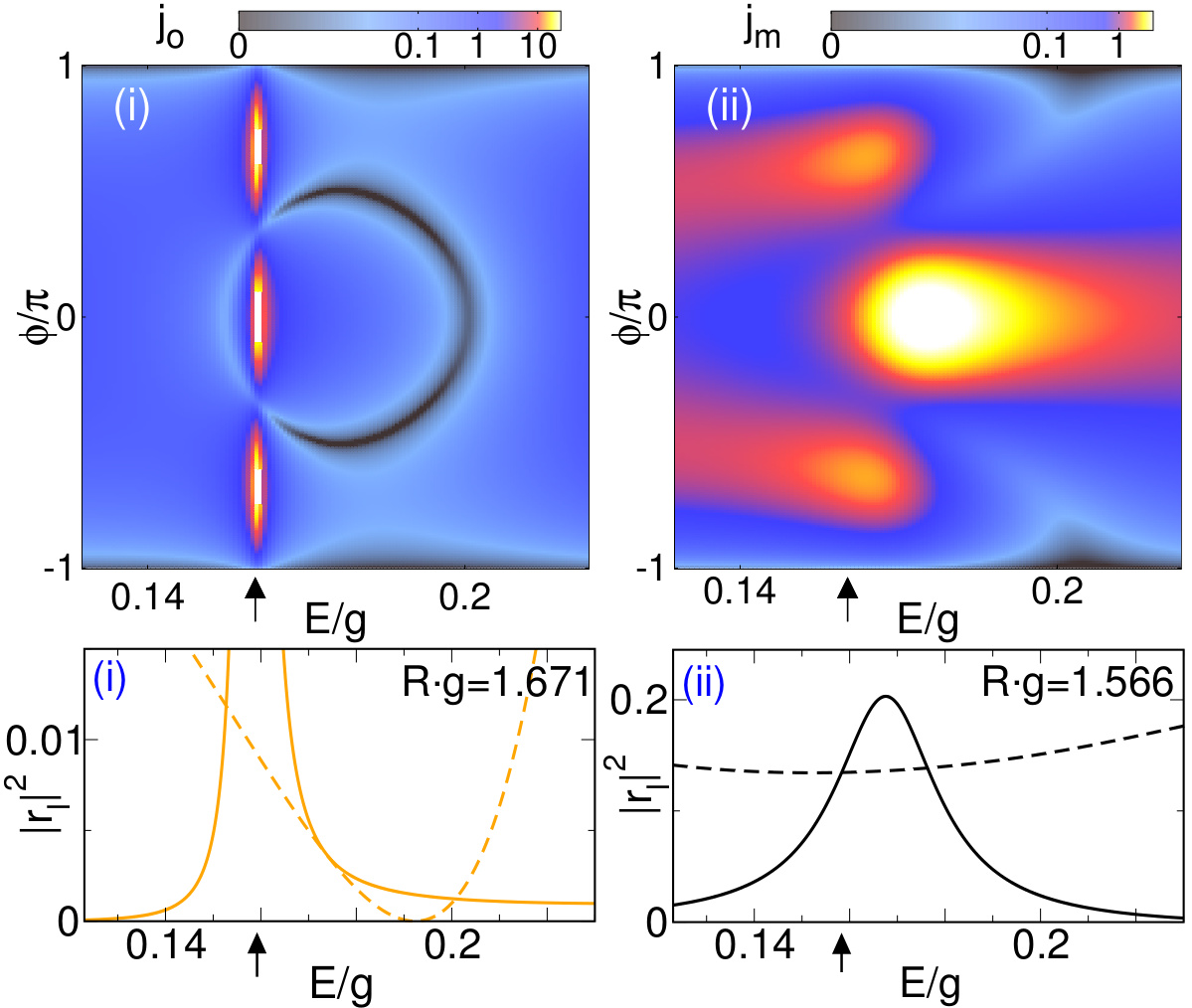}
      \caption{(Color online) Photonic ($j_{o}$) and phononic ($j_{m}$) angle-resolved far-field current [top] and first two photonic (orange) and phononic (black) reflection coefficients with $l=0$ (dashed) and $l=1$ (solid) [bottom] in dependence of $E/g$ for the cases (i) and (ii) in the middle panel of fig.~\ref{fig.3}. Again the phononic reflection coefficients $\absq{r_{m}}$ are divided by $v_{o}/4v_{m}$. Arrows mark the energy $E/g = 0.158$ used in the middle panel of fig.~\ref{fig.3}. \label{fig.5}}
\end{figure}

Having discussed the global scattering efficiency of the quantum dot, let us now analyze the spatial resolution of the wave transmisson and reflection. We start by investigating the scattering characteristics, specified by the probability density   
$\rho=\psi^{\dagger}\psi$ and current density $\bvec{j}=\psi^{\dagger}\hat{\bvec{j}}\psi$, in the near field, see fig.~\ref{fig.4}.
In the quantum dot region polaritons (mixed photon-phonon states) are formed.  For very small size-parameters $ER \ll 1$ and energies $E/g < 1$, the polariton density inside the dot becomes 
\begin{equation}\label{eq_density}
\absq{\psi^{t}}=(\absq{t_{+,l}}+\absq{t_{-,l}}) [J_{l}\bk{qr}^{2}+J_{l+1}\bk{qr}^{2}] 
-2\frac{v_{o}-v_{m}}{v_{o}+v_{m}}\mathfrak{Re}(t_{+,l}^{*}t_{-,l})t[J_{l}\bk{qr}^{2}-J_{l+1}\bk{qr}^{2}]\,.
\end{equation}
Obviously, $\rho$ is radially symmetric  (we have used that $q_{\pm} \to q=g/\sqrt{v_{o}v_{m}}$ for $E\to 0$). 
For resonant scattering the polariton density increases dramatically inside the dot, indicating a spatial and temporal `trapping' of
photon-phonon bound state, cf. fig.~\ref{fig.4}, left panels.  The resonance of the lowest partial wave $l=0$ confines the `quasiparticle'  about $r=0$, while resonances  with higher $l>0$ (not shown) give rise to ring-like structures close to the dot boundary related to `whispering gallery modes'. 

The current density inside the dot is given by
\begin{equation}\label{eq_current}
\bvec{j}^{t}=\frac{2v_{o}v_{m}}{v_{o}+v_{m}}(\absq{t_{+,l}}+\absq{t_{-,l}}) 
\times \{\cos\bk{\bk{2l+1}\phi}[J_{l+1}\bk{qr}^{2}+J_{l}\bk{qr}^{2}]\bvec{e}_{r} 
 + \sin\bk{\bk{2l+1}\phi}[J_{l+1}\bk{qr}^{2}-J_{l}\bk{qr}^{2}]\bvec{e}_{\phi}\}.
\end{equation} 
The panels right  in fig.~\ref{fig.4} show that the incident wave is fed into vortices which trap the polariton.  
For $l=0$, two vortices arise for the $n=1$ mode. Further vortices occur on the symmetry axis when $n$ increases.
In general, the vortex pattern of the $l$-th mode is dominated by $2(2l+1)$  vortices which give rise to $2l+1$ preferred scattering directions in the far field for $n=1$ (see below)~\cite{HBF13}.  We note that a very similar vortex pattern (scattering characteristics)  arises
for moderate size-parameters $ER \sim 1$, e.g., for the cases (i) and  (ii) in the middle panel of fig.~\ref{fig.3}.

The current density of the reflected waves in the far field given by eq.~\eqref{j} exhibits the already mentioned  cosinusoidal angle distribution with maxima at $\phi = l'\pi/\bk{2l+1}$ where $l' \in \left\{0,...,\pm l \right\}$. Consequently, if  the $l=0$ mode is resonant, only forward scattering takes place, whereas resonaces belonging to higher modes scatter the 
light respectively sound into different directions. This is illustrated by  fig.~\ref{fig.5} (upper panels), for the far-field currents $j_{o/m}$ 
of a specific quantum dot system that preferably suppresses either the phonon  [case(i)] or the photon  [case(ii)] scattering [cf. fig.~\ref{fig.3}, middle]. Accordingly, when the photonic partial wave with $l=1$ becomes resonant, we observe three preferred  scattering directions with equal intensity (left upper panel). Though a similar distribution results for the phononic resonance, now the forward scattering  is  somewhat enhanced as  the lower $l=0$ mode substantially contributes (right upper panel). Note that both waves will never be scattered in the angle range $\phi\simeq \pm \pi$ due to absence of backscattering. Most interestingly, the constructive and destructive interference between a resonant $l$ mode and the off-resonant $l=0$ mode can lead to a Fano resonance~\cite{F61} that for its part may cause a depletion of Klein tunneling, i.e., a suppression of  forward scattering~\cite{HBF13}. In this way, the interference between the first two photonic and phononic partial waves depicted in the lower panels of fig.~\ref{fig.5}  give rise to Fano resonances, which are reflected in the almost vanishing 
currents $j_{o/m}$ at certain ratios $E/g(\phi)$, even for $\phi=0$ (see upper panels). Varying the energy of the incident wave therefore allows to control the scattering into pure photon or phonon waves, having preferred directions of propagation, with or without forward scattering. 
\begin{figure}
\includegraphics[scale=0.261]{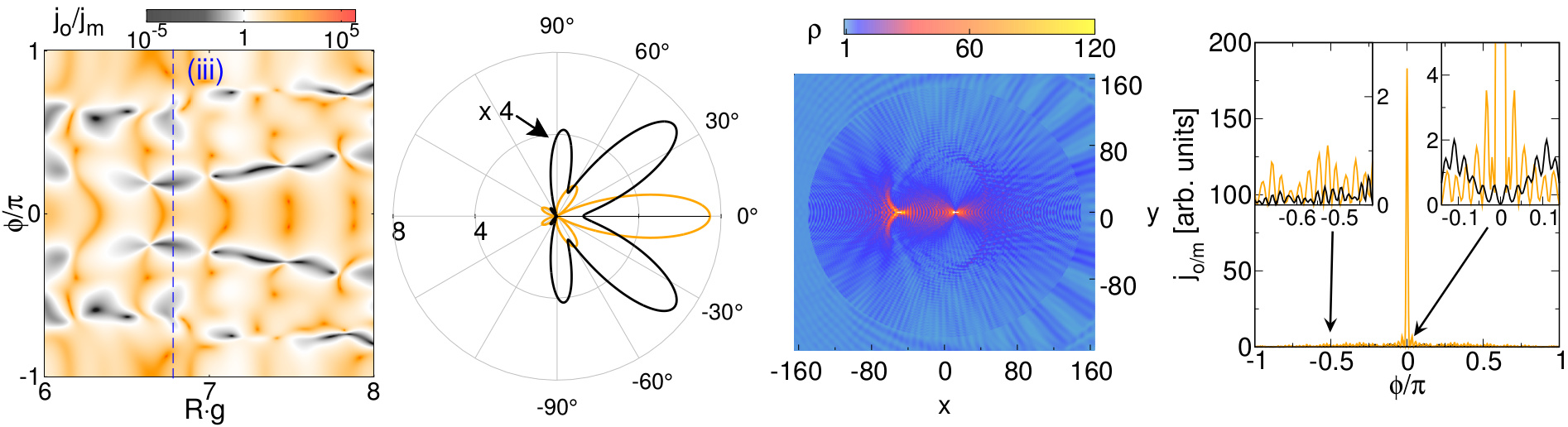}
\caption{(Color online) Left: angle-resolved ratio of photonic ($j_{o}$) and phononic ($j_{m}$) currents in the far field  depending on $Rg$. 
Middle left: polar plot of the photonic (orange) and phononic (black) far-field currents (arbitrary units) for case (iii) of fig.~\ref{fig.3} [right] (marked by the vertical blue dashed line in the left panel). 
The phononic current was multiplied by a factor of four. Middle right: probability density $\rho$ inside and outside the quantum dot. Right: photonic (orange) and phononic (black) currents in the far-field 
for $R=150$ ($g=1$),  $E=0.5$, i.e., the size-parameter $E R \gg 1$.\label{fig.6}}
\end{figure}

For larger size-parameters, $ER > 1$, where many partial waves may become resonant [e.g., case (iii) in fig.~\ref{fig.3} (right)], 
a rather complex structure of the far-field currents evolves. The two left panels in figure~\ref{fig.6}  display the ratio $j_o/j_m$ in the $Rg$--$\phi$ plane
and gives a polar plot of the light/sound emission. The figure corroborates the use of the considered setup as an optomechanical switch or light-sound translator. Finally, when $ER \gg 1$ and the extent of the quantum dot is much greater than the wavelengths, the scattering 
shows features known from ray optics [cf. fig.~\ref{fig.6}, middle right]. Such size parameters can only be realized by very large $R$, i.e., by a large number of cells (of the order of $10^{8}$) enclosed in the quantum dot region. The excitation of a large number of partial waves and their interference results in a caustics-like pattern of the transmitted wave inside the quantum dot and, most strikingly, the circular optomechanical barrier acts as a lens, focusing the light beam in forward direction, whereas the sound propagation is depleted [cf. fig.~\ref{fig.6}, right]. 
The far-field currents strongly oscillate when $\phi$ becomes finite, whereby the phonon contribution  is on average much smaller than those of the photon. 


To sum up, we have demonstrated Dirac physics in an optomechanical setting. Solving--within Dirac-Weyl theory--the problem of light scattering by circular barriers in artificial graphene composed of tunable optomechanical cells, we show that large quantum dots enable photon lensing, while small dots trigger the formation of polariton (photon-phonon) states which cause  a  spatial and temporal trapping of the incident wave in vortex-like structures, and a subsequent direction-dependent re-emittance of light and sound.  In the latter case (quantum regime), the quantum dot can be used to entangle photons and phonons and convert light to sound waves and vice versa.  Equally important, the forward scattering and Klein tunneling of photons could switched off for small dots by optically tuning a Fano resonance
arising from the interference between resonant scattering and the background partition. In this way optomechanical cells might be utilized to transfer, store, translate
and process information in (quantum) optical communications, or simply to realize a coherent interface between photons and phonons.

\InputIfFileExists{bib.bbl}

\section*{Acknowledgements}
This work was supported by Deutsche Forschungsgemeinschaft through SFB 652 (project B5). The authors wish to thank Andreas Alvermann for useful discussions at an early stage of this work.

\section*{Author contributions statement}
H. Fehske and C. Wurl outlined the scope of the paper and the strategy of the calculation. The calculations were performed by C. Wurl.
H. Fehske wrote the paper which was edited by C. Wurl. 

\section*{Additional information}
\textbf{Competing financial interests}
The authors declare no competing financial interests.



\end{document}